\documentstyle[12pt,aaspp4,flushrt]{article}
\def\etal{{\it et~al.~}}

\def\HST{{\it Hubble Space Telescope}}

\slugcomment{submitted to the Astrophysical Journal Letters}

\begin{document}

\title{GALAXY DARK MATTER: GALAXY-GALAXY LENSING 
IN THE {\sl HUBBLE DEEP FIELD}
\footnote{Based on observations with the NASA/ESA \HST, obtained at the
Space Telescope Science Institute, which is operated by AURA, under NASA
contract NAS 5-26555} 
}

\author{Ian P. Dell'Antonio and J. Anthony Tyson }
\affil{Bell Laboratories, Lucent Technologies, 700 Mountain Ave., 
Murray Hill, NJ 07974} 
\affil{Email: dellantonio@bell-labs.com, tyson@bell-labs.com}

\begin{abstract}

In this Letter we present calibrated observations of the average mass
of galaxies with $22< I < 25$ in the Hubble Deep Field.
We measure the mean mass profile using the
statistical gravitational lens distortion of faint
blue background galaxies, based on 2221 foreground-background pairs.
The observed weak lensing distortion 
is calibrated via full 3-D simulations and other HST data
on a foreground cluster of known mass.

Inside a projected radius of 5\arcsec\ we find a $3\sigma$ detection of
shear.
Fitting to an isothermal mass distribution,
we find a shear of 0.066$^{+0.018}_{-0.019}$ at a radius of 2\arcsec\ 
($\sim 8 ~ h^{-1}$ kpc).
The (1$\sigma$) limits on the parameters for the truncated isothermal model are 
$\sigma_v =185^{+30}_{-35}$ km s$^{-1}$, $r_{outer} \ge 15 ~ h^{-1}$ kpc, and
$r_{core}= 0.6^{+1.3}_{-0.6} ~ h^{-1}$ kpc. This corresponds 
to an average galaxy mass
interior to 20 $h^{-1}$ kpc of 
5.9$^{+2.5}_{-2.7}\times 10^{11}$ $h^{-1}$ M$_\odot$.
Inside 10 $h^{-1}$ kpc, we find an average rest-frame V band mass-to-light 
ratio of $11.4^{+6.0}_{-6.7}$ $h$ $(M/L_V)_\odot$.  

\end{abstract}


\section{Introduction}

The mass in the dark halos around individual field galaxies has long
been debated. While there are compelling kinematic data on our own
galaxy (Fich \& Tremaine 1991) and pairs of galaxies (Zaritsky and White 1994),
suggesting halos with equivalent circular velocities of $\sim$200 km sec$^{-1}$
extending as far as 30 kpc, direct measures of mass using weak gravitational 
lens distortions have been difficult. Typical galaxies produce $\sim 1\arcsec$
deflections, resulting in relatively small shear at radii $> 5\arcsec$.
Strong lensing of background QSOs has
provided a spot check of total enclosed mass in several cases 
(Kochanek 1991), confirming masses of 
$\sim 1 - 2 \times 10^{11}$ $h^{-1}$ M$_{\odot}$ 
within 4 $h^{-1}$ kpc.  
However, the known lensed QSOs may selectively sample the tail of the 
galaxy mass 
distribution (Wallington \& Narayan 1993).

Statistical gravitational lens distortion in a large sample of galaxies can
address the more general question of the mean mass of all galaxies.
Previous attempts to measure the average total mass
of a large sample galaxies by statistical weak lensing were
hampered by seeing and systematic error (Tyson, \etal 1984 [TVJM84]; 
Tyson, 1987).
Studying weak shear at radii $>5\arcsec$ in a 4.8\arcmin\ 
radius field taken with the Palomar 5-meter telescope, 
Brainerd \etal (1996) recently claimed evidence for massive halos
extending beyond 30 $h^{-1}$ kpc. 
The Hubble Deep Field (HDF) data, being
multi-band, high spatial resolution, and very deep -- despite the small size of
the field -- provide a large volume sample and an opportunity to control 
systematics and bring the $<$5\arcsec\ mass signal out of the noise.  

Here we report a calibrated measurement of the mass profile of field galaxies
in the HDF. 
Galaxies with the faint surface brightness and colors of the distant
faint blue galaxy population are used as background galaxies.
2221 foreground-background pairs of galaxies are examined for the 
induced shear from the foreground galaxy. We compare our result for the HDF
with full 3-D simulations, with deep HST weak lens studies around a cluster
of galaxies, 
and with broader but ``shallower" HST surveys.
A comprehensive description of the cluster HST WFC2 weak lens shear efficiency
calibrations, and comparison with other galaxy-galaxy weak lensing 
observations using HST and ground-based imaging will be given elsewhere.

\section{Image Processing and Photometry Catalogs} \label{proc}

For this study, we used the ``drizzled" HDF version 2 data in which the 
dithered images were combined on a resampling scale of 0.04\arcsec\ /pixel.  
The HDF observations and pipeline processing are described in 
Williams \etal 1996.  These reconstructed images cover 4.3 arcmin$^2$ 
for all three WFC2 chips.
The sky noise on 1\arcsec\ scales in these reconstructed images is
28 B mag arcsec$^{-2}$ and 28.5 R mag arcsec$^{-2}$.  
For this analysis we limit ourselves to the three deepest bands:  F814W, F606W, 
and F450W (hereafter called I, R, and B).  
We re-analyze the images using FOCAS to produce photometrically accurate
matched catalogs.  The zero-points in the various filter bands are set by the
HDF calibration (Williams \etal 1996).

We calculate magnitudes and
colors using an equal-isophote method:  initial object lists are constructed
by running FOCAS on B + R + I master images.
This produces a master list of objects and isophotal regions.
Magnitudes are then computed from the individual filter images using these
identical isophotal areas, so the color estimates are not biased
by systematic variations of the isophotes from band to band.  
This procedure yields magnitudes and intensity-weighted
moments measured accurately in three bands.
At a limiting magnitude of 29 B mag,  
about 40 galaxies between 24 and 29 mag will lie within a
10\arcsec\ circle projected on any foreground galaxy.

In order to extract the lensing shear, we must 
reliably measure the ellipticity and orientation of a faint galaxy.  
The faint background galaxies we use are quite small:
the profile of a typical 27th mag arclet is shown in Figure 1. Useful surface
photometry of these faint blue galaxies in the HDF data is possible only
within $r \sim 0.5$ arcsec.  Simulations of random
galaxy shapes using the observed noise characteristics of the HDF field
show that we can measure ellipticity for galaxies as faint as 
28 B mag and ellipticity $0.3$ with an uncertainty
in ellipticity less than $0.1$.  This suggests that we can safely use 
galaxies down to $B = 27.5$ as potential probes of the shear.  

\section{Arclet Analysis and Mass Model Simulations} \label{sim}

To minimize the contamination from foreground galaxies, we use the 
color and apparent magnitude of the galaxies as rough statistical 
redshift estimators.
Any galaxy in the magnitude range $22 < I < 25$ (roughly redshift 0.2 - 0.7) 
is a candidate
foreground lensing galaxy.  Blue galaxies with $B - R < 0.3$
that are fainter than a candidate foreground galaxy but brighter than 
27.5 B mag are candidate background galaxies. 
Due to competing volume and faint-end luminosity
function effects, these faint galaxies cover a wide range in
redshifts extending from z = 1--3; blue low surface brightness sources are
likely to be at the highest redshifts.  The trend towards redshift 1 
at 25th mag is clear in a magnitude -- log($z$) plot (Koo \& Kron 1992; 
Schade \etal 1995; Lilly \etal 1995).  

Using this selection, we find 110 candidate lenses and 645 candidate 
background objects yielding 2221 candidate foreground-background pairs 
in the three WFC2 CCD area of the HDF. 
We use FOCAS to calculate the intensity-weighted second moment
for each source galaxy image (arclet) within 15 arcsec of a candidate foreground
galaxy.  We calculate the $x^2$, $y^2$ and $xy$ intensity moments (apodized 
with a Gaussian kernel of size equal to the half-light radius)
transformed to $(r,\theta)$ moments relative to each foreground galaxy center
(TVJM84; Tyson, Valdes and Wenk 1990).
These principal-axis transformed second moments of the arclets 
within some annulus about all candidate foreground galaxies 
are averaged together to give the observed shear at that radius.

This observed shear as a function of radius for all 2221 pairs is plotted in
Figure 2. The points are the observed independent shear values in each radius 
bin, and the error 
bars obtained via the full simulations are consistent with
bootstrap resampling errors. As always, the largest noise source
is the randomly oriented non-zero ellipticities of the source galaxies.
We find a $3\sigma$ significance positive
mass effect observed at radii less than a few arcseconds
($\sim 10 ~ h^{-1}$ kpc for $0.2 < z_{fg} < 0.7$).  
While we have not ``tuned'' the arclet color to maximize shear, if we relax the 
blue selection this shear decreases; this is consistent with the blue B-R colors
of the highest redshift galaxies.

What galaxy mass distribution would generate this observed shear? In selecting
our sample galaxies based on brightness and color, the observed shear will be
diluted by the convolution of the separate redshift distributions 
of the candidate lenses and arclets.  The WFC2 point spread function (PSF)
and the drizzle algorithm have
an effect on observed shear for arclets
whose surface brightness falls below about 28 mag arcsec$^{-2}$ 
inside 0.2\arcsec\
radius. Therefore we must calibrate the lens mass which would generate this 
observed 
statistical shear. We do this in two steps: (1) full 3-D simulations 
of the HDF, 
and then (2) comparison of a similar 3-D simulation with the weak lensing shear 
of this same faint blue galaxy population observed with HST 
WFC2 around a cluster of galaxies of known redshift and mass.

\subsection{Mass calibration} \label{calib}

We calibrate the shear-mass relationship by simulating galaxy-galaxy lens 
distortions including multiple scattering effects.  Realistic simulated 
field galaxies consistent with mild luminosity evolution (Im \etal 1995)
are randomly distributed in seven redshift shells.  These simulated galaxies
have the same properties as 
the observed galaxies (number-magnitude, size-magnitude, 
ellipticity, and redshift distributions) as found in recent 
deep redshift and angular size surveys (Schade \etal 1995; Mutz \etal 1994). 
The redshift distribution we use is similar to that derived from four-color
photometry of the HDF (Mobasher \etal 1996), although lacking a distinct 
peak at $z\sim 2$.  

The lens distortions for galaxies in each redshift shell are computed using the 
galaxies in all of the shells in front of it.  After the cumulative distortions 
for all redshift shells are calculated, the image is convolved with 
the PSF, binned
down to the CCD resolution, and HDF-derived photon and readout noises are added.
All galaxies are modelled as soft core isothermal distributions
(Grossman \& Narayan 1988), but with an outer mass cutoff, giving 
for the spherical case a surface density distribution outside the core:
\begin{equation}
\Sigma = \Sigma_o \left({\beta}/2r\right)
	\left({1 + r^2/r^2_{outer}}\right)^{-1}
\end{equation}

\noindent
where $\Sigma_o = {{\sigma_v^2}/2G}$, $\sigma_v$ is the line-of-sight velocity
dispersion,
$r_{outer}$ is the outer mass cutoff, and 
$\beta = \left(1 + {{r_{core}^2}/{r_{outer}^2}}\right)$. 
For the candidate background galaxies, we measure the image distortion
in terms of the $(r,\theta)$ second moments $i_{rr}$ and $i_{\theta \theta}$:

\begin{equation}
T(r) = {i_{\theta \theta} - i_{rr} \over i_{\theta \theta} + i_{rr}} =
{\gamma(r) \over 1 - \kappa(r)}
\end{equation}
 
\noindent
where the convergence $\kappa(r)= \Sigma(r)/\Sigma_c$ 
and the shear $\gamma(r) = [\bar{\Sigma}(r) - \Sigma(r)]/\Sigma_c$ 
(Miralda-Escud\'{e} 1991),
and where $\Sigma_c$ is the critical surface mass density.
Outside the core, $\kappa << 1$, and we shall call $T(r)$ the ``shear".
We plot curves for the resulting simulated shear $T(r)$, corrected for the
cluster calibration and efficiency described below, for three values 
of foreground lens galaxy mass in Figure 2. 
All three have outer cutoff radii of $30 ~ h^{-1}$ kpc, 
soft core radii of 1.0$h^{-1}$ kpc, and line-of-sight 
velocity dispersions as labeled.  

The critical surface density is related to the distance
ratio:  $\Sigma_c=c^2/(4\pi G D)$.
The distance ratio for a foreground-background pair 
is (cf. Blandford and Narayan 1992)
\begin{equation}
D =  { (1 - q_o - d_1 d_2)(d_1 -d_2) \over 
(1 - q_o - d_2) (1 - d_2) (1 + z_{fg})} 
\end{equation}
where 
$d_1 = \sqrt{1+q_o z_{fg}}$ and $d_2 = \sqrt{1+q_o z_{bg}}$.  
By averaging over many different galaxies we are averaging $D$  over
the distributions of $z_{fg}$ and $z_{bg}$.
We combine these uncertainties, including multiple scattering
and the FOCAS arclet detection efficiency, in a single ``efficiency function"
$Q(r)$.  The value of $Q(r)$ is determined by comparing the 
prediction for single galaxy-galaxy lensing from equation (2) with the results 
of the 3-D simulations of the HDF galaxy-galaxy lensing, including
multiple scattering, normalizing to the galaxy cluster data.
In this way, we construct an absolute calibration for the shear-$\sigma_v$
relation.
A total of 4368 CPU-hours of time on an SGI Power Challenger 
were needed to perform the simulations.  Figure 3 shows a plot of $Q$ versus
$r$. 
Over the radial range of our data a constant $Q=0.33$ 
is a good overall approximation.  

If we have used the correct redshift distributions, etc., in our simulations,
this procedure should give reliable shear predictions.  We can largely
eliminate such redshift uncertainties by simulating a lens of known mass.
To fix the mass zero point for the simulation-based mass calibration we 
cross calibrate to a deep HST image of a massive cluster of 
galaxies at z = 0.4. The same 3-D simulation is run for
CL0024+1654 and compared with the observed weak shear in the HST WFC2 image of
this cluster, using identical arclet selection rules.
We find a similar (but 10 percent lower) efficiency for shear measurement in 
the cluster data at a radius where the mean shear is equal to the HDF shear
at $2\arcsec$.  The curves plotted in Figure 2 have these calibration
corrections applied. 

\subsection{Systematics}

A minor correction is made for the field distortion and the 
spatial variation of the WFC2 PSF 
by using star-dominated WFC2 images of the LMC.  Corrections due to the 
drizzle algorithm were found to be quite small:  measurement of the
signal using undrizzled data yielded statistically indistinguishable 
results.
Other systematic errors which would be present in the HDF data, 
via photometry
or sample selection or shear analysis, are present
also in the full simulations. Therefore, by comparing observed shear 
in the PSF-corrected
data and simulations, the derived limits on lens mass are not biased 
by systematics.  One sanity check is to
reverse the pairs: use arclets as lenses and lenses as arclets.
This gives a zero shear ($0.006\pm 0.019$). Another check
is to scramble the coordinates of arclets in the catalog and recalculate the
shear: again null ($0.009\pm 0.02$). Similarly, scrambling the positions of the
foreground galaxies yields a zero signal ($0.01\pm 0.02$).  Finally, 
each of the three WFC2 chips independently shows 
positive shear within 2\arcsec\ in the HDF.
As a check on the FOCAS software, we measured the shear signal from a set
of simulations identical to the ones used for calibration but with the 
lensing turned off.  This resulted in a null signal ($1.2\sigma$ negative
in the innermost bin).

However, spiral arms in foreground galaxies in the HDF might occasionally be
split off by FOCAS and mistaken for arcs. The simulations include
the ellipticity distribution of bright galaxies, but not spiral arms.
To check this, we take wider field deep ground-based imaging (0.5\arcsec\
pixels) and run FOCAS with the same parameters as with the HDF data, calling
each pixel 0.04\arcsec\ and scaling the magnitudes accordingly. This generates
38900 pairs of galaxies with scaled apparent magnitudes and angular 
sizes similar to the HDF. We obtain a null result for the observed
shear in this surrogate data: at 2\arcsec\ radius (25\arcsec\ in the 
original wide-field data)
we find a mean shear of $0.0011\pm 0.0016$, and at 1\arcsec\ we find 
$0.0019\pm 0.0025$. Thus, spiral arms are apparently not a
source of systematic error in the HDF shear measurements,
down to 1\arcsec\ radius, on the scale of our observed shear.

\section{Average Galaxy Mass} \label{mass}

Using the product of efficiency and average distance ratio $Q(r)$ found in the 
cluster-calibrated 3-D
simulations to normalize the analytic expression in equation 1 to the
HDF shear data, we find a best fit galaxy velocity dispersion $\sigma_v$
of 185$^{+30}_{-35}$ km sec$^{-1}$ for outer truncation radii larger
than 15 $h^{-1}$ kpc and $r_{core} \le 2 ~ h^{-1}$ kpc.  This corresponds
to a shear at 2\arcsec\ of $0.066^{+0.018}_{-0.019}$. 
The average mass interior to 20 $h^{-1}$ kpc is 
5.9$^{+2.5}_{-2.7}\times 10^{11}$ $h^{-1}$ M$_\odot$.
For $\sigma_v=185$ km sec$^{-1}$, the core radius is 
$r_{core}=0.6^{+1.3}_{-0.6} ~ h^{-1}$ kpc.  From the absence of an odd 
image in strongly lensed QSOs,
Wallington \& Narayan (1993) argue for mass cores $< 200$ pc.
The HDF data are consistent with this requirement, although our upper limits
are not as stringent.

Fixing $\sigma_v$ at 185 km sec$^{-1}$, we find
an acceptable fit for truncation radii 
$r_{outer} > 3.5 \arcsec$ ~ ($> 15 ~ h^{-1}$ kpc).
Inside $\approx 10 ~ h^{-1}$ kpc, 
the rest-frame V band mass-to-light ratio is 
$11.4^{+6.0}_{-6.7}$ $h ~ (M/L_V)_\odot$. 
VLA-based lensed QSO data sample smaller galaxy radii and find 
correspondingly smaller $M/L$
(see Fassnacht \etal 1996). Our $M/L$ for the whole field galaxy population
suggests a halo increasingly dominated by dark matter beyond 10 $h^{-1}$ kpc.
In this Letter we have only examined isothermal models.
The radial profiles of our observed and calculated shear are 
different; we will explore this at greater length elsewhere, combining various
deep HST and ground-based data. 

We briefly compare our results with the recent observations of galaxy-galaxy 
lensing
by Brainerd \etal (1996).  Although both studies probe galaxy-galaxy lensing,
they do so on different scales. 
The Brainerd \etal observations sample a region 
outside 5\arcsec\ or $\sim 15 ~ h^{-1}$
kpc for their sample.  By contrast, our sample shows strong shear precisely
inside this region. 
Extrapolating their shear at 10\arcsec\ to smaller radii yields a somewhat
greater shear than is seen in the HDF.
Ultimately, the distribution of galaxy mass on these larger scales will 
be addressed by comparably deep surveys covering hundreds of times the area 
of the HDF. 
In particular, these surveys will have to go significantly deeper than the
``Groth Strip" which covers 144 square arcmin, imaged with 
the HST in two filters: F555W and F814W.
This covers 27 times the area of the HDF, but 3 magnitudes shallower, 
resulting in fewer close pairs. Because the
candidate arclets are brighter and hence lower redshift, the 
lensing efficiency is also reduced.  For the Groth strip, we find 
a shear $0.015 \pm 0.009$ for 492 
foreground-background pairs within 5 arcseconds, roughly consistent with
the HDF results, although with much greater uncertainty.

We gratefully acknowledge the help of Tom Duff, Greg Kochanski and Rick Wenk, 
discussions with Andy Fruchter, Allen Mills, and Penny Sackett, and suggestions
of the referee. Phil Fischer kindly supplied the WFC2 LMC data.  
We thank Bob Williams and the HDF team
at STScI for rapidly making these data available to the community.

{}

\clearpage
\begin{figure}
\epsfysize = 6.2in \epsfbox{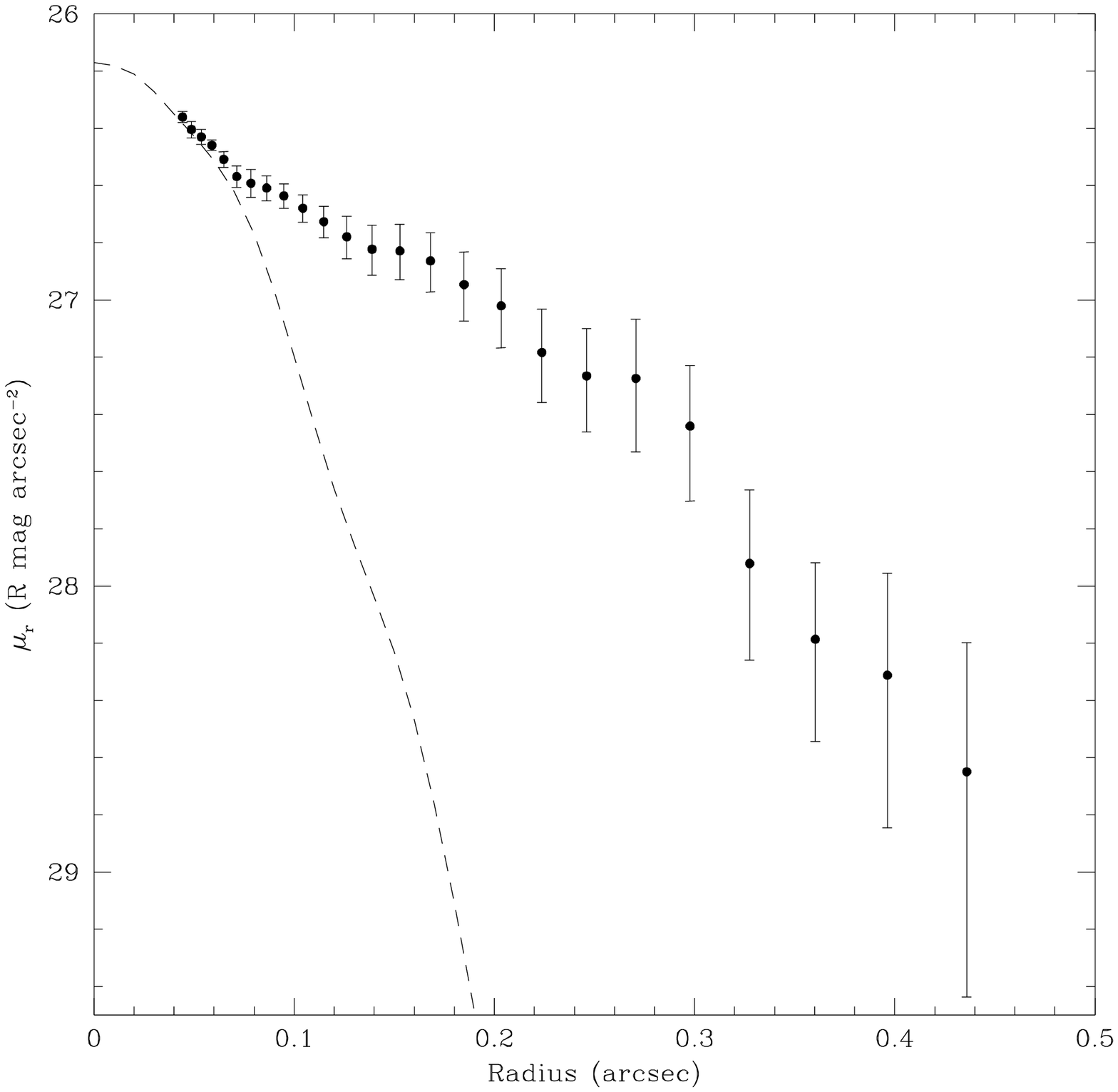}
\caption{Surface brightness profile of a typical 27 mag arclet
in the HDF. Accurate second moments for shear measurement require a limiting
surface brightness of 28-29 mag arcsec$^{-2}$ for these galaxies.  Note that
because the surface brightness is measured by averaging in elliptical annuli,
adjacent annuli sample some of the same pixels, and hence the data points
are correlated and the error bars only represent the photon statistics.
The dashed line is the average R-band WFPC2 PSF.}
\label{sbprof}
\end{figure}

\clearpage
\begin{figure}
\epsfysize = 6.6in \epsfbox{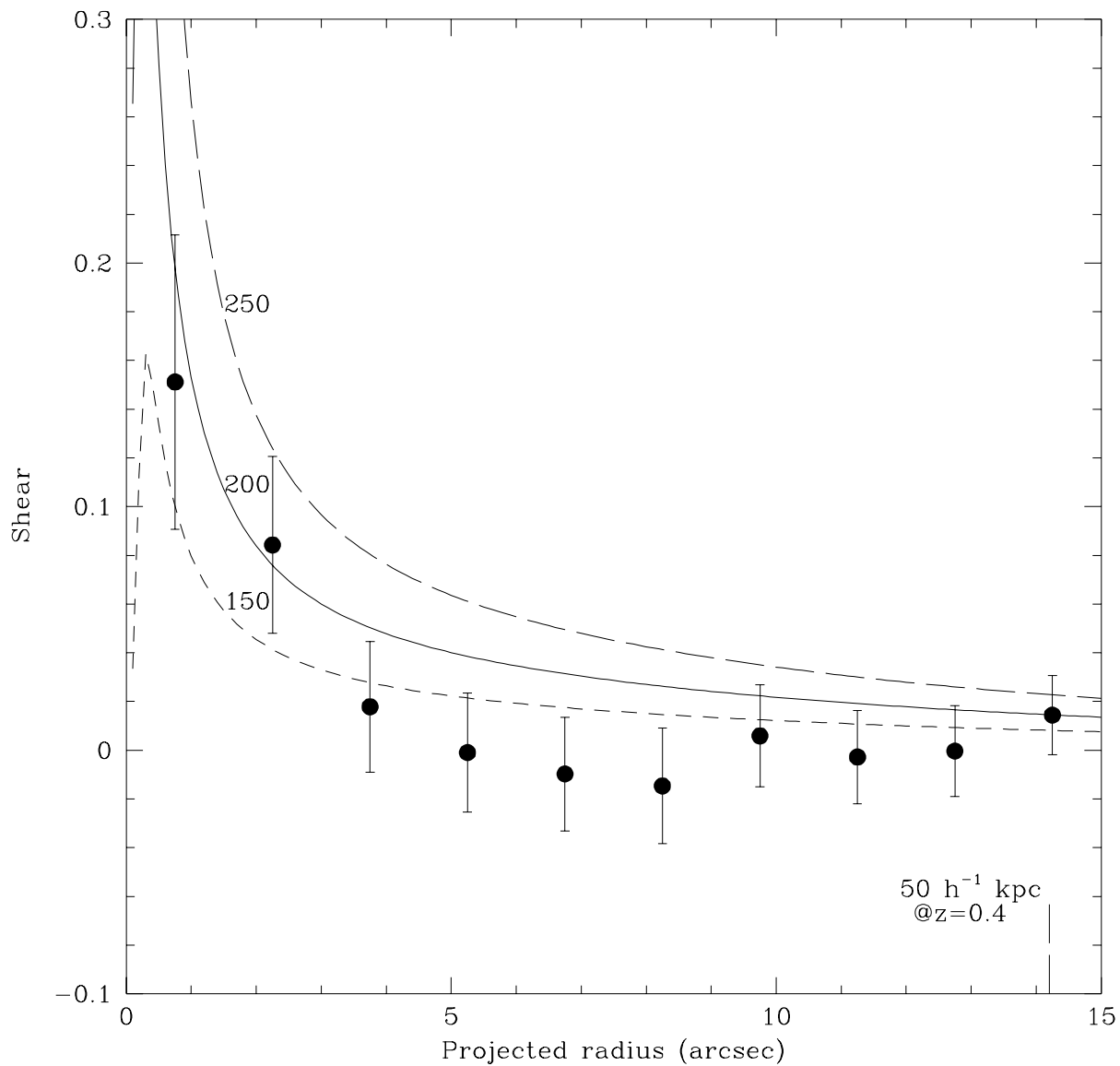}
\caption{The observed shear of the background galaxies, in
1.5\arcsec\ independent bins, around the positions of foreground galaxies
for 2221 foreground-background pair candidates in the three WFC fields of the
HDF.
The background/foreground selection was based on colors, surface brightness,
and total magnitude difference.
The lines show shear from full 3-D calibrated 
simulations of truncated isothermal models with $\sigma_v$ as labeled,
$r_{outer} = 30 ~ h^{-1}$ kpc, and $r_{core}=1 ~ h^{-1}$ kpc.
}
\label{shear}
\end{figure}

\clearpage
\begin{figure}
\epsfysize = 6.2in \epsfbox{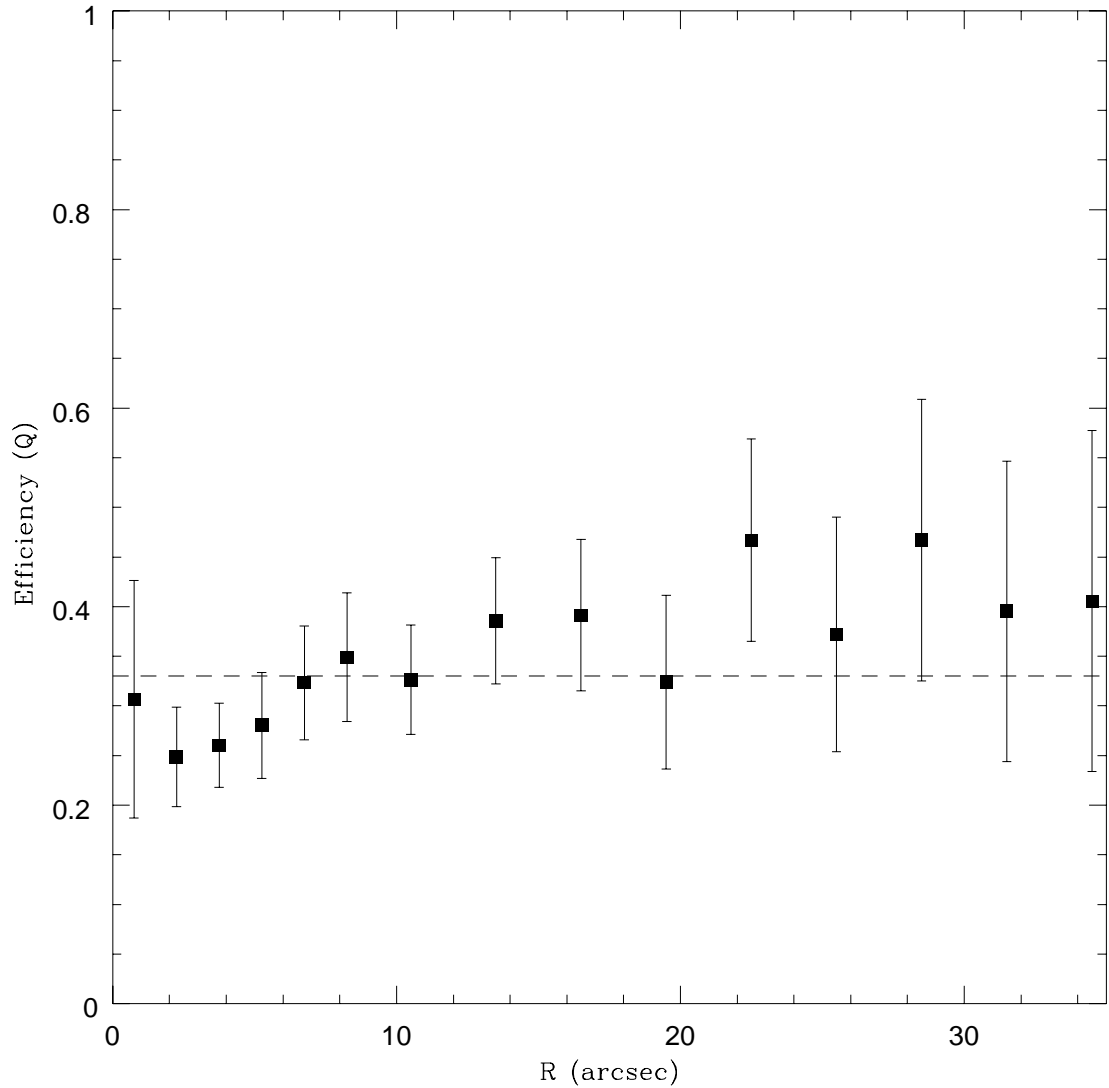}
\caption{The product of distance ratio and efficiency for recovery of 
shear as a function of radius from the foreground galaxy,
from the simulations. $Q$ includes the effects of source and lens 
redshifts, and is calibrated by normalizing $Q_{sim}/Q_{obs}$ to a lens of 
known mass, CL0024+1654.  The $1\sigma$ error bars are derived 
from the scatter in measured shear in the simulations. The dashed line
is the value of Q used in plotting the curves in Figure 2.}
\label{qplot}
\end{figure}

\end{document}